# The Relationship between Tsallis Statistics, the Fourier Transform, and Nonlinear Coupling

Kenric P. Nelson and Sabir Umarov

*Abstract* — Tsallis statistics (or *q*-statistics) in nonextensive statistical mechanics is a one-parameter description of correlated states. In this paper we use a translated entropic index: $1\text{-}q \to q$. The essence of this translation is to improve the mathematical symmetry of the *q*-algebra and make *q* directly proportional to the nonlinear coupling. A conjugate transformation is defined $\hat{q} \equiv \frac{-2q}{2+q}$ which provides a dual mapping between the heavy-tail *q*-Gaussian distributions, whose translated *q* parameter is between $-2 < q < 0$, and the compact-support *q*-Gaussians, between $0 < q < \infty$. This conjugate transformation is used to extend the definition of the *q*-Fourier transform to the domain of compact support. A conjugate *q*-Fourier transform is proposed which transforms a *q*-Gaussian into a conjugate $\hat{q}$-Gaussian, which has the same exponential decay as the Fourier transform of a power-law function. The nonlinear statistical coupling is defined such that the conjugate pair of *q*-Gaussians have equal strength but either couple (compact-support) or decouple (heavy-tail) the statistical states. Many of the nonextensive entropy applications can be shown to have physical parameters proportional to the nonlinear statistical coupling.

## I. INTRODUCTION

THE developments in the field of nonextensive entropy [1] have had a significant impact on analytical methods for modeling statistical behavior of nonlinear systems. The physics of nonextensive systems have been shown to provide important analytical techniques for analysis of turbulence [2], communication signals [3, 4], dynamics of the solar wind [5], neural networks [6] and other complex phenomena.

In this paper the expressions for *q*-statistics [7] are simplified and connected more directly to nonlinear analysis, by a simple translation of the *q*-parameter. As currently expressed, the nonlinearity inherent in the *q*-Gaussian and other *q*-functions disappears when $q = 1$. This definition originates from the raising of probabilities to the power *q* [8], but it does not utilize the symmetry of the real numbers about zero.

Moreover, the physical interpretation of what the *q* parameter represents has remained obscure. These conceptual difficulties are eliminated by a simple translation. In the expressions for *q*-Gaussians and *q*-algebra, the term $(1-q)$ appears frequently. By an explicit translation of the current parameter, either $q' \to 1-q$ or equivalently $1-q' \to q$, the *q*-algebra is made symmetrical with the real numbers and the physical significance of *q*-statistics can be related directly to the nonlinear coupling strength. This convention is used in [9] and will be expanded upon here. In the remainder of the paper the original parameter of *q*-statistics is denoted by $q'$ to distinguish it from the proposed translation $q$.

This simple, and by first appearances unremarkable translation, has non-trivial consequences. First, the generalization is now centered about the natural symmetry of the real numbers, so for $q = 0$, the *q*-Gaussian is the familiar bell-shaped Gaussian. This symmetry about zero makes it possible to express the fundamental equations of *q*-statistics and *q*-algebra without burdensome numerical constants. The structure of the simplified *q*-algebra is introduced in Section II and detailed in Appendix A. In section III the mathematical symmetry is utilized to define a conjugate dual between heavy-tail and compact-support distributions. This important relationship is used to extend and improve the definition of the *q*-Fourier transform in Sections IV and V. Most significantly, the proposed variable *q* emerges as a direct measure of the nonlinear coupling in the statistical system, which is discussed in Section VI. Thus the simplification of the *q*-statistic equations leads to symmetric mathematical relationships and a stronger connection to physical principles.

K. P. Nelson is with Raytheon Company, Woburn, MA 01801 USA (corresponding author 603-508-9827; kenric.nelson@ieee.org).

S. Umarov., is with Tufts University, Medford, MA 02155 USA (sabir.umarov@tufts.edu).

## II. Preliminaries

Tsallis [8] showed that raising probabilities by a power leads to a nonextensive generalization of the Boltzmann-Gibbs-Shannon entropy function relevant to the thermodynamic properties of nonlinear systems. Further developments in this field have led to a $q$-algebra [10], which encapsulates the nonlinear relationships in $q$-statistics in a form which provides analogs to basic mathematical relationships. The building block for the $q$-algebra is the $q$-exponential function and the $q$-addition which defines how the exponents of $q$-exponentials are combined. However, the current expression for these two functions suffers from the need to include $1-q$ in the definition. As currently expressed, the $q$-exponential is

$$e_{q'}^x \equiv (1+(1-q')x)_+^{1/(1-q')} \qquad (1)$$
$$= \begin{cases} (1+(1-q')x)^{1/(1-q')} & 1+(1-q')x \geq 0 \\ 0 & 1+(1-q')x < 0 \end{cases}$$

and two exponents combined by $q$-addition is $x \oplus_{q'} y = x + y + xy - q'xy$. Scaling the $q$-addition and the other functions of $q$-algebra to many variables is complicated by the accumulation of numerical constants. However, by translating $1-q'$ to $q$, the $q$-addition not only is easier to express, but makes explicit that $q$ is the nonlinear coupling between the variables $x$ and $y$. Now $q$-addition is $x \oplus_q y = x + y + qxy$ and the nonlinear coupling is zero at $q = 0$. The $q$-exponential function simplifies to

$$e_q^x \equiv (1+qx)_+^{1/q} = \begin{cases} (1+qx)^{1/q} & 1+qx \geq 0 \\ 0 & 1+qx < 0 \end{cases} \qquad (2)$$

For $\lim_{q \to 0} e_q^x = e^x$ and there is no nonlinear coupling, but as $q$ increases so does the coupling between the variables. Negative values of q are associated with a 'decoupling' or anti-correlation between the variables. A full description of the translated $q$-algebra is provided in appendix A.

A simple illustration of the utility of the $q$-exponential function is that it is the solution for the nonlinear differential equation [8]

$$y^{1-q} = -\frac{dy}{dt} \text{ or } y = -y^q \frac{dy}{dt}. \qquad (3)$$

The solution to this equation is $y = e_q^{-t} = (1-qt)^{1/q}$. For $q = 0$, the differential equation is linear and its solution is the exponential decay function. For $q > 0$ the nonlinear coupling increases the rate of decay. Instead of decaying to zero as $t \to \infty$, the solution "decays" to zero at the finite value of $t = 1/q$. This faster-than-exponential decay is referred to as 'compact-support' if the negative values beyond the zero-root are truncated to the value zero. For $q < 0$ the nonlinear coupling decreases the rate of decay. The solution now decays at a slower-than-exponential rate, referred to as a 'heavy-tail'.

The nonextensive entropy function provides a starting point for interpreting the proposed translation of the $q$-parameter. Entropy is the average of the 'surprisal', $\ln(1/p_i)$:

$S = \langle \ln(\frac{1}{p_i}) \rangle = -\sum_{i=1}^{n} p_i \ln p_i$. Following from the $q$-exponential, the logarithm can be generalized to a function of fractional power by $\ln_q(x) \equiv \frac{x^q - 1}{q}$, where $q$ is the translated parameter of $q$-statistics, and the natural logarithm is recovered as $q$ approaches zero. Using the generalized logarithm, the nonextensive entropy is defined as

$$S_q \equiv \langle \ln_q(1/p_i) \rangle = \sum_{i=1}^{n} p_i \left( \frac{p_i^{-q} - 1}{q} \right) = \frac{-1 + \sum_{i=1}^{n} p_i^{1-q}}{q}. \qquad (4)$$

With the translated nonextensive entropy function, the power of the probability is $1-q$. The one represents the probability required for the average, and the negative sign is due to the 'surprisal' being the inverse of the probability, and the nonlinear coupling term $q$ follows from the generalization of the logarithm. The defining expression for nonextensive entropy, which relates the $q$-entropy for two independent systems $A$ and $B$, now has the nonlinear term scaled by $q$

$$S_q(A+B) = S_q(A) + S_q(B) + qS_q(A)S_q(B) \qquad (5)$$

Using the escort probability with a power of $1-q$ as a constraint leads to the $q$-Gaussian distribution as the maximum $q$-entropy solution [11]. The escort probability is defined as

$$P_i = \frac{p_i^{1-q}}{\sum_{i=1}^{n} p_i^{1-q}}. \qquad (6)$$

Further insight into this form of the escort probability is possible by multiplying the numerator and denominator by all of the non-zero probabilities raised to the $q$-power

$$P_{q,i} = \frac{p_i/p_i^q}{\sum_{i=1}^{n} p_i/p_i^q} = \frac{p_i \prod_{\substack{j=1 \\ j \neq i}}^{m} p_j^q}{\sum_{i=1}^{n} p_i \prod_{\substack{j=1 \\ j \neq i}}^{m} p_j^q}, \qquad (7)$$

where $n$ is the number of states and $m$ is the number of states with non-zero probability. Here the escort probability is expressed to show that the statistical properties of the i$^{th}$ state of the nonlinear system include the nonlinear coupling via $q$ to all the other states of the system with non-zero probability. The numerator is expanded to further illustrate this expression

$$p_i [p_1 \cdot p_2 \cdots p_{i-1} \cdot p_{i+1} \cdots p_m]^q. \qquad (8)$$

In this form, the numerator of the escort probability explicitly includes the nonlinear coupling between all the states of a system. The denominator normalizes this expression. Because of the significance of nonlinear coupling and its explicit expression in this probability distribution, the raising by the power 1-q and renormalization of a probability distribution will be referred to as *the coupled probability*

*distribution.*

In continuous variables, the coupled probability density function and the generalized entropy are

$$f_q(x) = \frac{[f(x)]^{1-q}}{\int_{-\infty}^{+\infty} [f(x)]^{1-q} dx} \quad (9)$$

$$S_q = \frac{-1 + \int_{-\infty}^{\infty} f^{1-q}(x) dx}{q}. \quad (10)$$

The q-mean and q-variance of a random variable X are defined using the coupled probability density function

$$\mu_q = \int_{-\infty}^{\infty} x f_q(x) dx$$
$$\sigma_q^2 = \langle (X - \mu_q)^2 \rangle_q = \int_{-\infty}^{\infty} (x - \mu_q)^2 f_q(x) dx. \quad (11)$$

Reference [12] includes a development of the all the q-moments for arbitrary densities. The maximum of the q-entropy with finite q-variance is the q-Gaussian

$$G_q(x) = \frac{\sqrt{\beta_q}}{C_q}\left[1 - q\beta_q(x - \bar{\mu}_q)^2\right]_+^{\frac{1}{q}} \equiv \frac{\sqrt{\beta_q}}{C_q} e_q^{-\beta_q(x-\bar{\mu}_q)^2} \quad (12)$$

where $e_q^x$ is the q-exponential as defined in (2), $\beta_q = [(2+q)\sigma_q^2]^{-1}$ and the normalization term $C_q$ is

$$C_q = \begin{cases} \sqrt{\frac{\pi}{q}} \frac{\Gamma\left(\frac{1+q}{q}\right)}{\Gamma\left(\frac{2+3q}{2q}\right)} & \Leftarrow \quad q > 0 \\ \sqrt{\pi} & \Leftarrow \quad q = 0 \\ \sqrt{\frac{\pi}{-q}} \frac{\Gamma\left(\frac{2+q}{-2q}\right)}{\Gamma\left(\frac{1}{-q}\right)} & \Leftarrow \quad -2 < q < 0 \end{cases} \quad (13)$$

Thus, the q-Gaussian has its origin in the nonlinear coupling of statistical states, as defined in the coupled probability density and the generalized entropy. For positive values of q the nonlinear coupling between the states strengthens the decay, resulting in a distribution with compact support. In this case, the probability is zero for $|x - \mu_q| > \sqrt{1/q\beta_q}$. At q=0 there is no nonlinear statistical coupling and the traditional Gaussian distribution, common throughout linear systems analysis, holds. For negative values of q the decoupled states result in weakening of the decay, resulting in a heavy-tail distribution. Beyond $q < -\frac{2}{3}$ the classic variance is divergent, but the q-variance is finite. Beyond $q < -2$ the distribution can not be normalized; i.e. $C_q$ is divergent.

The arguments of the gamma function within the normalization term are related to a sequence of q values which were defined as a property of repeated application of the q-Fourier transform [13]. Here we show that the q-sequence is also related to the derivative and integral of q-Gaussians. The derivative of the q-exponential is

$$\frac{d}{dx}\left[e_q^{ax}\right] = \frac{d}{dx}\left[(1+qax)^{\frac{1}{q}}\right] = a(1+qax)^{\frac{1}{q}-1} \quad q \neq 0. \quad (14)$$

The power $\frac{1}{q} - 1 = \frac{1-q}{q}$ is equivalent to the q-parameter transitioning from $q \to \frac{q}{1-q}$. Expressing the derivative with the q-exponential notation and extending to the $n^{th}$-derivative results in

$$\frac{d}{dx}\left[e_q^{ax}\right] = a\left(1 + \frac{q}{1-q}(1-q)ax\right)_+^{\frac{1-q}{q}} = a\exp_{\frac{q}{1-q}}[(1-q)ax], \quad q \neq 1$$
$$\frac{d^n}{dx^n}\left[e_q^{ax}\right] = \left[a^n \prod_{i=1}^{n}(1-(i-1)q)\right]\exp_{\frac{q}{1-nq}}[(1-nq)ax], \quad q \neq \frac{1}{n}. \quad (15)$$

The integral of the q-exponential has the following form

$$\int e_q^{ax} dx = \frac{1}{a(1+q)}\left(1 + \frac{q}{1+q}(1+q)ax\right)^{\frac{1+q}{q}}$$
$$= \frac{1}{a(1+q)}\exp_{\frac{q}{1+q}}[(1+q)ax] + c_1, \quad q \neq -1$$
$$\int\ldots\left(\int e_q^{ax} dx\right)\ldots dx = \left[\frac{1}{a^n}\prod_{i=1}^{n}\frac{1}{1+nq}\right]\exp_{\frac{q}{1+nq}}[(1+nq)ax] \quad (16)$$
$$+ \sum_{i=1}^{n} c_i x^{i-1}, \quad q \neq -\frac{1}{n}.$$

The expressions for the derivative and integral of the q-Gaussian are complicated by the $x^2$ term in the exponent. Rather than develop the full form, it is sufficient for the current discussion to note that the $n^{th}$ integral (excluding the integer polynomial) and the $n^{th}$ derivative of the q-Gaussian have the following relationship

$$\text{power of } n^{th} \text{ q-Gaussian derivative: } \tfrac{2}{q} \to \tfrac{2}{q} - n$$
$$\text{power of } n^{th} \text{ q-Gaussian integral: } \tfrac{2}{q} \to \tfrac{2}{q} + n \quad (17)$$

These q-values are the same as those for the $n^{th}$ q-Fourier Transform. Summarizing, the $n^{th}$ q-parameter is defined as

$$z_n(q) \equiv q_n \equiv \frac{2q}{2+nq} = \left[\frac{1}{q} + \frac{n}{2}\right]^{-1} \quad n = 0, \pm 1, \pm 2, \ldots \quad (18)$$

Positive values of n are related to the $n^{th}$ integral of the q-Gaussian. Negative values of n are related to the $n^{th}$ derivative. The numeral 2 is related to the power of 2 for the Gaussian distribution and generalizes to $0 < \alpha \leq 2$ for the $(q, \alpha)$- distributions [6, 8-10]

$$G_{q,\alpha}(x) \equiv ae_q^{-\beta|x|^\alpha} = a\left(1 - q\beta|x|^\alpha\right)^{1/q} \quad (19)$$

$$z_{(a,n)} \equiv q_{(a,n)} \equiv \frac{\alpha q}{\alpha + nq} = \left[\frac{1}{q} + \frac{n}{\alpha}\right]^{-1} \quad (20)$$
$$0 < \alpha \leq 2; \; n = 0, \pm 1, \pm 2, \ldots$$

The $(q, \alpha)$-distribution is part of the generalization of the alpha-stable Lévy distributions. Using the translated q-parameter the q-sequence can be expressed without numerical constants. The $q_{(\alpha,n)}$ term will be abbreviated by the equivalent expression $q_{2n/\alpha}$, since $q_{(\alpha,n)} = q_{2n/\alpha} = \frac{2q}{2+\left(\frac{2n}{\alpha}\right)q}$. For double-sided q-exponential functions, $\alpha = 1$; for Gaussian functions $\alpha = 2$. The ability to express this important relationship in a simple intuitive form is in sharp contrast to

the original expression

$$q'_{(\alpha,n)} = \frac{n+(\alpha+n)q'}{\alpha+n(1-q')} \qquad (21)$$

The clarity of the translated expression for the $q$-sequence simplifies the establishment of many important relationships within the $q$-algebra. For example the normalizing coefficient for the $q$-Gaussian $C_q$ can be expressed as

$$C_q = \begin{cases} \sqrt{\frac{\pi}{q}} \frac{\Gamma\left(\frac{1}{q_2}\right)}{\Gamma\left(\frac{1}{q_3}\right)} & \Leftarrow \quad q > 0 \\ \sqrt{\pi} & \Leftarrow \quad q = 0 \\ \sqrt{\frac{\pi}{-q}} \frac{\Gamma\left(\frac{1}{-q_1}\right)}{\Gamma\left(\frac{1}{-q}\right)} & \Leftarrow \quad -2 < q < 0 \end{cases} \qquad (22)$$

The normalization can be simplified in other ways by applying the relationship $\Gamma(x+1) = x\Gamma(x)$: $\Gamma(\frac{1}{q_2}) = (\frac{1}{q}+1) = \frac{1}{q}\Gamma(\frac{1}{q})$ and $\Gamma(\frac{1}{q_3}) = \Gamma(\frac{1}{q_1}+1) = \frac{1}{q_1}\Gamma(\frac{1}{q_1})$.

### III. CONJUGATE PAIRS OF HEAVY-TAIL AND COMPACT-SUPPORT Q-GAUSSIANS

The $q$-sequence can be used to define a mapping between the heavy-tail $q$-Gaussians, with $-2 < q < 0$ and the compact-support $q$-Gaussians, with $0 < q < \infty$. For guidance in this mapping, first consider the Fourier transform for a power-law function, which the heavy-tail q-Gaussians approach asymptotically as x goes to infinity. For $-2 < q < 0$, the tail of the q-Gaussian approaches a power-law

$$e_q^{-\beta x^2} \sim (-q\beta x^2)^{\frac{1}{q}} \sim O(x^{\frac{2}{q}}), \quad x \to \infty \qquad (23)$$

The Fourier transform of a power law is also a power law with the following form [14, 15]

$$|x|^{\frac{2}{q}} \Leftrightarrow \sqrt{\frac{2}{\pi}}\Gamma(\frac{2}{q}+1)\sin(-\frac{\pi}{q})|\omega|^{-(\frac{2}{q}+1)} \quad \text{for } -2 < q < 0 \qquad (24)$$

Where $\omega$ is the dual variable (or frequency). In the frequency domain the power is $-(\frac{2}{q}+1)$ which rearranged in the form of a q-Gaussian, shows the following relationship with the q-sequence

$$|w^2|^{-\frac{2+q}{2q}} = |w^2|^{-\frac{1}{q_1}} \qquad (25)$$

So the Fourier transform of a power law is suggestive of an important mapping between a q-Gaussian and a $-q_1$-Gaussian.

In fact the mapping between $G_q \leftrightarrow G_{-q_1}$ constitutes a conjugate dual relating the heavy-tail $q$-Gaussians to the compact-support $q$-Gaussians. Figure 1 shows $-q_1$ as a function of $q$. The compact-support range $q > 0$ is mapped to the heavy-tail range $-2 < -q_1 < 0$ and vice-versa. This relationship leads to the definition for the *q-conjugate dual*.

*Definition 1* The $q$-conjugate dual is defined as

$$\hat{q} \equiv -z_1(q) = \frac{-2q}{2+q}. \qquad (26)$$

The inverse of the conjugate is the same function:

$\hat{q}^{-1} = \frac{-2\hat{q}}{2+\hat{q}} = \frac{-2(-q_1)}{2+(-q_1)} = q$. The $q$-conjugate of a function is determined by applying the $q$-conjugate to each of the $q$ parameters of the function. However, care must be taken if the conjugate is applied to the value $\pm q_k$. A broader hat will be used to clarify that the conjugate includes the sign and the sequence value.

*Definition 2* The q-conjugate for $\pm q_k$ and the inverse are

$$\widehat{\pm q_k} \equiv \frac{-2(\pm q_k)}{2+(\pm q_k)} = \mp q_{k\pm 1};$$
$$\widehat{\mp q_{k\pm 1}} = \frac{-2(\mp q_{k\pm 1})}{2+(\mp q_{k\pm 1})} = \pm q_k. \qquad (27)$$

This notation is required to contrast with the $n^{th}$ sequence of a $q$-conjugate. The $n^{th}$ sequence value of a $q$-conjugate is

$$\pm \hat{q}_k = \pm z_k(\hat{q}) = \pm \frac{2(-q_1)}{2+k(-q_1)} = \mp q_{1-k} \qquad (28)$$

When applying the conjugate to a $q$-function care is needed to apply the conjugate to the same $q$-sequence value throughout the function.

*Definition 3:* Let $X_{q_k}$ be the set of functions which depend on $q_k$. Then the q-conjugate operator and its inverse are defined as

$$T_{(q,\hat{q})}: X_{q_k} \to X_{\hat{q}_k}$$
$$T_{(\hat{q},q)} = T_{(q,\hat{q})}^{-1} = T_{(q,\hat{q})}. \qquad (29)$$

For $f(q_k;x) \in X_{q_k}$; $\tilde{f}(q_k;x) \equiv T_{(q,\hat{q})}(f(q_k;x)) = f(\tilde{q}_k;x)$.

A set of conjugate $q$-Gaussian pairs with $\sigma_q^2 = \sigma_{\hat{q}}^2 = 1$ is shown in Figure 2. On the left of Figure 2 are the compact-support distributions which proceed from the Gaussian distribution for values close to zero and converge to the uniform distribution as $q$ goes to infinity. On the right are the conjugate heavy-tail distributions, which converge to a uniform distribution of infinite extent and infinitesimal density at $q = -2$. If $\beta$ is invariant the compact-support distribution converges to a delta function as $q$ goes to infinity.

The integral of the unnormalized $q$-Gaussian and $\hat{q}$-Gaussian functions have the following relationship. Restating the definitions

$$\int_{-\infty}^{\infty} e_q^{-\beta_q x^2} dx = C_q / \sqrt{\beta_q} = \sqrt{2+q}\sigma_q C_q$$
$$\int_{-\infty}^{\infty} e_{\hat{q}}^{-\beta_{\hat{q}} x^2} dx = C_{\hat{q}} / \sqrt{\beta_{\hat{q}}} = \sqrt{2+\hat{q}}\sigma_{\hat{q}} C_{\hat{q}} \qquad (30)$$

If $-2 < q < 0$, then $0 < \hat{q} < \infty$ and the normalization constant $C_{\hat{q}}$ is determined from (22)

$$C_{\hat{q}} = \sqrt{\frac{\pi}{-q_1}} \frac{\Gamma\left(-\frac{1}{q_{-1}}\right)}{\Gamma\left(-\frac{1}{q_{-2}}\right)} = \sqrt{\frac{\pi}{-q_1}} \frac{\left(-\frac{1}{q_1}\right)\Gamma\left(-\frac{1}{q_1}\right)}{\left(-\frac{1}{q}\right)\Gamma\left(-\frac{1}{q}\right)} \qquad (31)$$

The ratio between $C_{\hat{q}}$ and $C_q$ is

$$\frac{C_{\hat{q}}}{C_q} = \frac{\sqrt{\frac{\pi}{-q_1}}\left(\frac{q}{q_1}\right)\frac{\Gamma\left(-\frac{1}{q_1}\right)}{\Gamma\left(-\frac{1}{\hat{q}}\right)}}{\sqrt{\frac{\pi}{-q}}\frac{\Gamma\left(-\frac{1}{q_1}\right)}{\Gamma\left(-\frac{1}{q}\right)}} = \left(\frac{q}{q_1}\right)^{3/2} = \left(\frac{2+q}{2}\right)^{3/2} \quad (32)$$

This ratio also holds if $0 < q < \infty$ and is thus definitive of the integrals of complimentary heavy-tail and compact-support $q$-Gaussian functions. Figure 3 shows both normalization constants as a function of their respective q values and their ratio. The normalization terms in (30) are equal if the variance of the $\hat{q}$-Gaussian function is

$$\sigma_{\hat{q}}^2 = \left(\frac{2+q}{2-q_1}\right)\left(\frac{C_q}{C_{\hat{q}}}\right)^2 \sigma_q^2 = \frac{(2q/q_1)}{(2q/q)}\left(\frac{q_1}{q}\right)^3 \sigma_q^2 = \left(\frac{q_1}{q}\right)\sigma_q^2. \quad (33)$$

If the normalization is invariant then the conjugate pairs have equal density at $x = 0$.

An instructive illustration of the how the conjugate pairs strengthen the interpretation of $q$-Gaussians, is the relationship with the Student-T distribution and the $\kappa$-distribution. The Student-T [16, 17] originates from random variable $T = \frac{Z}{\sqrt{V/\nu}}$ where Z has the standard normal distribution and V has a chi-squared distribution with $\nu$ degrees of freedom. The distribution is $f(t) = \frac{\Gamma(\frac{\nu+1}{2})}{\sqrt{\nu\pi}\Gamma(\frac{\nu}{2})}\left(1+\frac{t^2}{\nu}\right)^{-(\frac{\nu+1}{2})}$; which can be equated to a heavy-tail $q$-Gaussian $\frac{\sqrt{\beta}}{C_q}e_q^{-\beta t^2}$ with $q = -\left(\frac{2}{\nu+1}\right)$ and $\beta = \left(\frac{\nu+1}{2\nu}\right) = \frac{1}{2+q}$, which is equivalent to $\sigma_q^2 = 1$. The complement of $q$ is $\hat{q} = \frac{2}{\nu}$, which demonstrates an inverse relationship between the $q$-conjugate and the degree of freedom. In Section VI the interpretation of physical applications with heavy-tail distributions will also be simplified by utilizing the $q$-conjugate parameter.

Leubner and Voros [18, 19] have discussed the relationship between Tsallis statistics and the $\kappa$-distribution, used by the space physics community to model power-law distributions in plasma velocities. Both approaches have the same form, and as defined by Leubner and Voros have the simple translation $\kappa = 1/q$ using the proposed $q$ parameter or $\kappa = 1/(1-q')$ using the original parameter. The $q$-sequence transitions defined by (18) are easier to express, $\kappa_n = \kappa + n/2$, since $\kappa$ represents the power of the generalized exponential directly rather than the inverse. However, the exponential and Gaussian functions are recovered as $\kappa \to \infty$, which complicates the relationship between these fundamental functions and there generalization to 'compact-support' and 'heavy-tail' distributions.

In the next two sections the conjugate pairs of $q$ parameters are used to extend and improve the nonlinear generalization of the Fourier transform.

IV. THE $q$-FOURIER TRANSFORM FOR COMPACT-SUPPORT DOMAIN

In contrast to the Fourier transform where the power of the $q$-Gaussian is transformed from $\frac{2}{q} \to -\frac{2}{q_1}$, but the form of the function is no longer a $q$-Gaussian, the $q$-Fourier transform [13, 20, 21] preserves the $q$-Gaussian form with the power shifting from $\frac{2}{q} \to \frac{2}{q_1}$. Previously, the $q$-Fourier transform was only defined in the region of heavy-tail distributions, $-2 < q < 0$, because the Cauchy Integral Theorem used in the definition is not applicable for the compact-support region. The definition can now be extended to the compact-support domain by utilizing the conjugate dual. The extension consists of converting the compact-support distributions to the conjugate heavy-tail distributions, performing the $q$-Fourier transform, and then converting the heavy-tail solution to the compact-support solution. The result is identical in form to that already demonstrated for the heavy-tail distributions.

*Definition 4*: Let $-2 < q \leq 0$. The q-Fourier transform is defined for a nonnegative function $f(x) \in L_1(R)$ using the q-exponential and the q-product

$$F_q[f](\omega) = \int_{\text{supp } f} e_q^{ix\omega} \otimes_q f(x)dx = \int_{-\infty}^{\infty} f(x)e_q^{ix\omega[f(x)]^{-q}}dx. \quad (34)$$

*Definition 5*: Let $q > 0$. The q-Fourier transform for the compact-support domain is defined as

$$\begin{aligned}F_q[f](\omega) &= T_{(\hat{q},q)}\left[\int_{\text{supp } f} T_{(q,\hat{q})}[e_q^{ix\omega} \otimes_q f(x)]dx\right] \\ &= T_{(\hat{q},q)}\left[\int_{-\infty}^{\infty} T_{(q,\hat{q})}[f(x)e_q^{ix\omega[f(x)]^{-q}}]dx\right].\end{aligned} \quad (35)$$

*Lemma 6:* The functional form of $F_q[f](\omega)$ is equal for the compact-support domain ($q > 0$) and the heavy-tail domain $-2 \leq q < \infty$.

*Proof:* Since

$$T_{(q,\hat{q})}[e_q^{ix\omega} \otimes_q f(x)] = e_{\hat{q}}^{ix\omega} \otimes_{\hat{q}} f(x) \text{ and}$$

$$\int_{\text{supp } f} e_{\hat{q}}^{ix\omega} \otimes_{\hat{q}} f(x)dx = F_{\hat{q}}[f](w),$$

then $F_q[f](\omega) = T_{(\hat{q},q)}[F_{\hat{q}}[f](\omega)]$ and Definition 4 and Definition 5 have the same functional form. □

*Corollary 7:* Let $q > -2$. For the $q$-Gaussian $G_q(x) = ae_q^{-\beta x^2}$ the $q$-Fourier Transform is

$$F_q\left[G_q(x)\right](w) = Ae_{q_1}^{-Bx^2}, A = \frac{aC_q}{\sqrt{\beta}}, B = \frac{(2+q)}{8\beta a^{2q}}. \quad (36)$$

This relationship is derived in Example 9.

*Example 8*: As an illustration of the $q$-Fourier transform applied to a general function, consider the transformation of the uniform distribution as a function of $q$. The uniform distribution is defined as

$$U(x) = \begin{cases} \frac{1}{2} & |x| \leq 1 \\ 0 & \text{elsewhere} \end{cases}. \quad (37)$$

For $-2 < q \leq 0$ the $q$-Fourier transform is

$$F_q[U(x)](\omega) = \int_{-\infty}^{\infty} U(x) e_q^{ixw[U(x)]^{-q}} dx$$

$$= \int_{-1}^{1} \frac{1}{2}[1 + qixw(\frac{1}{2})^{-q}]^{\frac{1}{q}} dx$$

$$= \frac{1}{2qiw2^q} \int_{-1}^{1} [1 + qixw2^q]^{\frac{1}{q}} (qiw2^q) dx \quad (38)$$

$$= \frac{q/_{1+q}}{(2i)wq2^q} \left[ (1 + qiw2^q)^{\frac{1}{q}+1} - (1 - qiw2^q)^{\frac{1}{q}+1} \right]$$

$$= \frac{1}{(2i)(1+q)2^q w} \left[ e_{q_2}^{i(1+q)2^q w} - e_{q_2}^{-i(1+q)2^q w} \right]$$

$$= \frac{\sin_{q_2}[(1+q)2^q w]}{(1+q)2^q w} = \text{sinc}_{q_2}[(1+q)2^q w].$$

The solution makes use of the $q$-sin [22], which is defined as $\sin_q(x) = \frac{e_q^{ix} - e_q^{-ix}}{2i}$; and the $q$-sinc, which is defined as $\sin_q(w)/w$. The properties of $\text{sinc}_{q_2}[(1+q)2^q w]$ are examined in Figure 4. For $q = 0$, the solution is the sinc function which is consistent with the Fourier transform. For negative values of $q$ the oscillations of the $q$-sin and $q$-sinc functions are dampened. Between $-\frac{1}{3} < q < 0$, the dampening is soft enough that there is a least one transition through zero. Between $-1 < q < -\frac{1}{3}$, the oscillations are over-damped and the function decays to zero at infinity without any negative values.

The same solution can be derived for $q > 0$ using the conjugate transformation. In this case in accordance with (35)

$$F_q[U(x)](\omega) = T_{(\hat{q},q)} \circ \int_{-\infty}^{\infty} U(x) T_{(q,\hat{q})} [e_q^{ixw[U(x)]^{-q}}] dx$$

$$= T_{(\hat{q},q)} \circ \int_{-1}^{1} \frac{1}{2}[1 + \hat{q}ixw(\frac{1}{2})^{-\hat{q}}]^{\frac{1}{\hat{q}}} dx$$

$$= T_{(\hat{q},q)} \circ \frac{\hat{q}/_{1+\hat{q}}}{(2i)w\hat{q}2^{\hat{q}}} \left[ (1 + \hat{q}iw2^{\hat{q}})^{\frac{1}{\hat{q}}+1} - (1 - \hat{q}iw2^{\hat{q}})^{\frac{1}{\hat{q}}+1} \right] \quad (39)$$

$$= \frac{1}{(2i)(1+q)2^q w} \left[ e_{q_2}^{i(1+q)2^q w} - e_{q_2}^{-i(1+q)2^q w} \right]$$

$$= \frac{\sin_{q_2}[(1+q)2^q w]}{(1+q)2^q w} = \text{sinc}_{q_2}[(1+q)2^q w].$$

The solution makes use of the transformation from $\hat{q}_2$ to $q_2$, defined in (28). Figure 4c shows the q-Fourier transform for the uniform distribution for several values between $0.01 \leq q \leq 1$. In this region, the oscillations of the q-sinc function are amplified. A logarithmic plot is used to show the changing scale of the oscillations. Eventually the amplification overwhelms the oscillations. Near $q = 0.5$ the function no longer resembles a sinc function and is negative except for values of $w$ near zero. At $q = 1$ the function is one for all frequencies. Although not shown, for values of $q > 1$ the amplification is dominant and the function approaches infinity without any oscillations.

*Example 9 Derivation of Corollary 7:* If $f(x) = ae_q^{-\beta x^2}$ and is heavy-tail, $-2 < q < 0$, the $q$-Fourier transform is shown in [13] to be

$$F_q\left[ ae_q^{-\beta x^2} \right](w) = \left( \frac{aC_q}{\sqrt{\beta}} e_q^{-\frac{w^2}{4\beta a^{2q}}} \right)^{\frac{2+q}{2}}$$

$$= Ae_{q_1}^{-Bw^2}; \quad q_1 = \frac{2q}{2+q}; \quad A = \frac{aC_q}{\sqrt{\beta}} \quad B = \frac{(2+q)}{8\beta a^{2q}}. \quad (40)$$

Expressing $\beta$ in terms on the $q$-variance, $\beta = 1/_{(2+q)\sigma_q^2}$, the exponent is $B = \frac{(2+q)^2 \sigma_q^2}{8a^{2q}} = \left( \frac{(2+q)\sigma_q}{2\sqrt{2}a^q} \right)^2$. Making use of the relationship between compact-support and heavy-tail $q$-Gaussians, the $q$-Fourier transform for $q > 0$, can be shown to have the same form.

Starting with a compact-support $q$-Gaussian function, $f(x) = ae_q^{-\beta x}$, $q > 0$, the conjugate heavy-tail function is determined using conjugate mapping $q \to \hat{q}$. For purposes of defining the $q$-Fourier transform for compact-support $q$-Gaussians, the constants $a$ and $\beta$ are treated as independent of $q$. The definition still holds if either of these constants is a function of $q$ and is thus transformed. The conjugate $q$-function is then $\hat{f}(x) = T_{(q,\hat{q})} \circ f(x) = \hat{a}e_{\hat{q}}^{-\hat{\beta}x}$, $-2 < \hat{q} < 0$, $\hat{q} = -q_1$, with $\hat{a} = a$, $\hat{\beta} = \beta$ because they are treated as independent of $q$ for this discussion. The $q$-Fourier transform of this distribution is

$$F_{\hat{q}}\left[ \hat{f}(x) \right](w) = (a \frac{C_{\hat{q}}}{\sqrt{\beta}} e_{\hat{q}}^{-\frac{w^2}{4\beta a^{2\hat{q}}}})^{\frac{2+\hat{q}}{2}}$$

$$= \hat{A} e_{\hat{q}_1}^{-\hat{B}w^2}; \quad \hat{q}_1 = \frac{2(-q_1)}{2+(-q_1)} = -q; \quad (41)$$

$$\hat{A} = a \frac{C_{\hat{q}}}{\sqrt{\beta}}; \quad \hat{B} = \frac{(2+\hat{q})}{8\beta a^{2\hat{q}}}.$$

The function $-q$-Gaussian is transformed to $q_1$-Gaussian by repeating the transformation $T$ this time with the parameters $-q = \hat{q}_1 \to q_1$. Applying $T$ to (41) results in

$$F_q[f(x)](w) = T \circ F_{\hat{q}} \circ T[f(x)](w) = T\left[ \hat{A} e_{\hat{q}_1}^{-\hat{B}w^2} \right]$$

$$= Ae_{q_1}^{-Bw^2}, \quad A = a\frac{C_q}{\sqrt{\beta}}, B = \frac{(2+q)}{8\beta a^{2q}} \quad (42)$$

which is identical to (40), the result for heavy-tail $q$-Gaussians. Thus, the definition for the $q$-Fourier transform is shown to be consistent for the full range of $q$-Gaussian distributions, $q > -2$.

*Definition 10*: Let $q > -2$. The $q$-Fourier transform is defined for $f(x) \in L_1(R)$ as

$$F_q[f(x)](w) = \begin{cases} \int_{\text{supp } f} e_q^{ixw} \otimes_q f(x) dx & -2 < q \leq 0, \\ T_{(\hat{q},q)} \circ \int_{\text{supp } f} e_{\hat{q}}^{ixw} \otimes_{\hat{q}} \hat{f}(x) dx & q > 0 \end{cases} \quad (43)$$

$$\hat{f}(x) = T_{(q,\hat{q})}[f(x)]$$

For simplicity of expression, the conjugate transformation of the integrand in the domain $q > 0$, is implied by specify $\hat{q}$

as the nonlinear parameter. Following the integration, the conjugate transformation is specified directly. In the original $q$-algebra notation the extended definition of the $q$-Fourier transform has the following form.

*Definition 11*: Let $q' < 3$. The $q'$-Fourier transform is defined for $f(x) \in L_1(R)$ as

$$F_{q'}[f(x)](w) = \begin{cases} \int_{supp\, f} e_{q'}^{ixw} \otimes_{q'} f(x)dx & 1 \le q' < 3, \\ T_{(\hat{q}',q')} \circ \int_{supp\, f} e_{\hat{q}'}^{ixw} \otimes_{\hat{q}'} \hat{f}(x)dx & q' < 1 \end{cases} \quad (44)$$

$$\hat{f}(x) = T_{(q',\hat{q}')}[f(x)]; \quad \hat{q}' = \tfrac{5-3q'}{3-q'}$$

An important consequence of the extension of the $q$-FT to include the domain of compact is its implications for the $q$-Central Limit Theorem ($q$-CLT). This generalization of the Central Limit Theorem utilized the $q$-product and $q$-Fourier transform to prove that distributions coupled via the $q$-product converged in the limit to a $q$-Gaussian distribution. The $q$-CLT can now be extended to the compact-support domain.

V. DEFINITION OF THE CONJUGATE $q$-FOURIER TRANSFORM

The relationship between heavy-tail and compact-support $q$-Gaussians is suggestive of a *conjugate q-Fourier transform*, which will be closer in function to the actual Fourier transform. In this alternative form, the $q$-Gaussian is transformed to a $\hat{q}$-Gaussian which approximates the actual Fourier transform. An additional benefit is that all finite-mean $q$-Gaussian distributions have a finite-mean conjugate $q$-Fourier transform, in contrast to the $q$-FT which requires $q > -1$ for the transformation to have $q > -2$.

A requirement for the transformation to result in a $\hat{q}$-Gaussian is for the $q$ parameter to be transformed to $-q_2$ prior to applying the $q$-FT. The transformation between $q$ and $-q_2$ is also a conjugate dual, but is related to the heavy-tail and compact-support 2-sided exponential distributions.

*Definition 12:* Let $\alpha = 1$ for the $(\alpha, q)$-exponential family. The $q$-exponential conjugate dual is defined using (20) as

$$\tilde{q} \equiv -z_{(1,1)}(q) = -q_2 = \frac{-q}{1+q}$$
$$\tilde{q}^{-1} = \tilde{\tilde{q}} = \frac{-\tilde{q}}{1+\tilde{q}} = q. \quad (45)$$

*Definition 13*: The q-exponential conjugate and its inverse for the set of q-functions ($X_{q_k}$) is

$$T_{(q,\tilde{q})} : X_{q_k} \to X_{\tilde{q}_k}$$
$$T_{(\tilde{q},q)} = T_{(q,\tilde{q})}^{-1} = T_{(q,\tilde{q})}. \quad (46)$$

For $f(q_k; x) \in X_{q_k}$; $\tilde{f}(q_k; x) \equiv T_{(q,\tilde{q})}(f(q_k; x)) = f(\tilde{q}_k; x)$.

*Definition 14 (The heavy-tail $\tilde{q}-FT$):* Let $\tilde{q}$ be in the heavy-tail $q$-Gaussian domain $-2 < \tilde{q} \le 0$. Then the conjugate $q$-Fourier transform ($\tilde{q}$-FT or $F_{\tilde{q}}[f](w)$) is defined as

$$F_{\tilde{q}}[f(x)](w) = \int_{supp\, f} e_{\tilde{q}}^{ixw} \otimes_{\tilde{q}} \tilde{f}(x)dx \quad -2 < \tilde{q} \le 0 \quad . \quad (47)$$

*Definition 15 (The compact-support $\tilde{q}-FT$):* Let $\tilde{q}$ be in the compact-support $q$-Gaussian domain $\tilde{q} > 0$. Then the conjugate $q$-Fourier transform is defined as

$$F_{\tilde{q}}[f(x)](w) = T_{(\hat{\tilde{q}},\tilde{q})} \circ \int_{supp\, f} e_{\hat{\tilde{q}}}^{ixw} \otimes_{\hat{\tilde{q}}} \hat{\tilde{f}}(x)dx \quad \tilde{q} > 0,$$
$$\hat{\tilde{q}} = \frac{-2\tilde{q}}{2+\tilde{q}} = \frac{-2(-q_2)}{2+(-q_2)} = q_1. \quad (48)$$

*Lemma 16:* The conjugate $q$-FT of $f(x)$ is equal to the $q$-exponential conjugate of the $q$-FT

$$\tilde{q}\text{-}FT[f](w) = T_{(q,\tilde{q})}[q\text{-}FT[f](w)].$$

*Proof:* For $-2 < \tilde{q} \le 0$,

$$T_{(q,\tilde{q})}[q\text{-}FT[f](w)] = T_{(q,\tilde{q})}\left[\int_{supp\, f} e_{q}^{ixw} \otimes_{q} f(x)dx\right]$$

which is equivalent to (47). For $\tilde{q} > 0$ the solution has same form from *Lemma 6*.

*Example 17*: Reexamining the two examples from the previous section, the uniform distribution $U(x)$ and the general $q$-Gaussian function, have conjugate $q$-Fourier transforms of the following form. Applying *Lemma 16* the solution is simply the $q$-exponential conjugate of the original $q$-FT. The $\tilde{q}-FT$ for the uniform distribution is

$$F_{\tilde{q}}[U(x)](w) = \text{sinc}_{\tilde{q}_2}[(1+\tilde{q})2^{\tilde{q}} w]$$
$$= \text{sinc}_{-q}[(1-q_2)2^{-q_2} w] \quad (49)$$

The properties of the solution graphed in Figure 4 are still relevant, but the 'compact-support' and 'heavy-tail' regions are swapped. Thus for $q > 0$, the solution is now a $-q$-sinc function with damped oscillations; and for $q < 0$ the $-q$-sinc function has amplified oscillations.

*Example 18:* Likewise, the $\tilde{q}$-FT for a q-Gaussian is

$$F_{\tilde{q}}\left[ae_q^{-\beta x^2}\right](w) = \left(\tfrac{a\tilde{C}_{\tilde{q}}}{\sqrt{\beta}} e_{\tilde{q}}^{-\tfrac{w^2}{4\beta_{\tilde{q}} a^{2\tilde{q}}}}\right)^{\tfrac{2+\tilde{q}}{2}}$$
$$= \tilde{A} e_{\tilde{q}_1}^{-\tilde{B} w^2}; \quad \tilde{q}_1 = \tfrac{2\tilde{q}}{2+\tilde{q}} = -q_1 = \hat{q}; \quad (50)$$
$$\tilde{A} = \tfrac{a\tilde{C}_q}{\sqrt{\beta}}; \quad \tilde{B} = \tfrac{(2+\tilde{q})}{8\beta a^{2\tilde{q}}} = \tfrac{(2-q_2)}{8\beta a^{-2q_2}}.$$

Again, the coefficients $a$ and $\beta$ are treated as independent of $q$. The solution is a $\hat{q}$-Gaussian with coefficients A and B are functions of $\tilde{q} = -q_2$. Thus for compact-support $q$-Gaussian with $q > 0$, the $\tilde{q}$-FT is a heavy-tail $q$-Gaussian with $-2 < \hat{q} < 0$; and vice-versa for the 'heavy-tail' $q$-Gaussians. This transformation has two attractive properties a) all $q$-Gaussians in the range $-2 < q < 0$ have a well-defined conjugate $q$-Fourier transform and b) the fundamental property of the Fourier transform whereby 'wide' signals transform to a 'narrow' frequency domain with the Gaussian distribution as the symmetric invariant is preserved.

## VI. THE RELATIONSHIP BETWEEN $q$ AND NONLINEAR STATISTICAL COUPLING

The conjugate pair of $q$-parameters provides a key insight into the relationship between Tsallis Statistics and the coupling strength of nonlinear systems. This section will provide an interpretation of the applications discussed in [1] and other references, which shows that for applications with 'compact-support' $q$-Gaussian distributions $q$ is proportional to the nonlinear coupling, and that applications with 'heavy-tail' $q$-Gaussian distributions the conjugate $\hat{q}$ is proportional to the nonlinear coupling. Table 1 provides a synopsis of this interpretation. As a demonstration of the relationship between Tsallis statistics and nonlinear coupling, we will review an application of particular relevance to information systems, multiplicative noise.

Drawing upon the analysis by Anteneodo and Tsallis [23], consider a stochastic process influenced by multiplicative and additive Gaussian white noise

$$dX_t = f(X_t) + g(X_t)(2MdW_t^{(m)}) + (2AdW_t^{(a)}) \quad (51)$$

where $X(t)$ is the stochastic variable, $f$ is the deterministic drift and $g$ is the noise-induced drift, and $dW_t^{(m,a)}$ are independent Wiener processes which define the multiplicative ($m$) and additive ($a$) noise, and $M$ and $A$ are the amplitude of the noise. The probability density $p_X(x,t)$ for this system is determined using the Focker-Planck equation:

$$\frac{\partial p_X(x,t)}{\partial t} = -\frac{\partial}{\partial x}[J(x)p_X(x,t)] + \frac{\partial^2}{\partial x^2}[D(x)p_X(x,t)] \quad (52)$$

where $J(x)$ is the drift and $D(x)$ is the diffusion:

$$J(x) \equiv f(x) + Mg(x)g'(x)$$
$$D(x) \equiv A + M[g(x)]^2 \quad (53)$$

If the deterministic and noise-induced drift are related by a potential $V(x) = \frac{\tau}{2}[g(x)]^2$, then $f(x) = -V'(x) = -\tau g(x)g'(x)$. The drift term simplifies to

$$J(x) = f(x)(1 - \frac{M}{\tau}) \quad (54)$$

which clarifies that the deviation from deterministic drift induced by the multiplicative noise has a strength of $M/\tau$.

In this case, the stationary solution $p_X(x)$ for (52) is a $q$-Gaussian

$$p_X(x) = \lim_{t \to 0} p_X(x,t) \propto e_q^{-\beta[g(x)]^2}; \quad \beta = \frac{\tau + M}{2A}; \quad q = \frac{-2M}{\tau + M} \quad (55)$$

The $q$-conjugate transformation (29) is $\hat{q} = \frac{2M}{\tau}$, which demonstrates that the $\hat{q}$-parameter is directly proportional to the nonlinearity of the system $M/\tau$. As Table 1 shows this simple, intuitive relationship is consistent for many of the nonextensive entropy applications. Some analytical models, such as the nonlinear Fokker-Planck equation, in which the parameter (in this case $v$) is valid for all real values do not follow the pattern just described. The thermodynamics of system in a finite bath described by Anrade, et. al. [24] is an example of $q$-exponential model in which case q is equal to the nonlinear coupling. In this example $q = 1/\beta H$, where $\beta$ is the inverse temperature and $H$ is the total energy.

The applications examined here lead to a conjecture regarding *nonlinear statistical coupling*. Whereas, the nonlinear coupling defined by $q$ from the $q$-algebra expressions can be any real number, the nonlinear statistical coupling is constrained by the requirement of a finite-mean distribution for $q > -\alpha$. Nevertheless, it is conjectured that the nonlinear statistical coupling strength is the positive real numbers for both the heavy-tail and compact-support regions. For the heavy-tail region the strength of the nonlinear statistical coupling is determined using the conjugate relationship.

*Conjecture 19:* Let $\phi$ be the nonlinear statistical coupling strength which results in the $(q,\alpha)$-distribution being characteristic of a statistical properties of the system. Then $\phi$ is related to the entropic index q and the power of the q-exponential variable α by the following relationship

$$\phi = \begin{cases} q/\alpha & q \geq 0 \\ \dfrac{-q}{\alpha + q} & -\alpha < q < 0 \end{cases} \quad (56)$$

*Remark 20*: Issues which need to be examined to *evaluate whether* Conjecture 19 is valid include: a) a precise definition of the nonlinear statistical coupling strength for a variety of physical systems. b) determination of when the physical parameters of a nonlinear stochastic process are constrained to lead to $q > -\alpha$ b) physical examples with $\alpha \neq 1, 2$ and a more detailed description of the $q$-generalized $\alpha$-stable distributions c) when a nonlinear stochastic process has solutions with divergent mean $q < -\alpha$ what is the relationship between $q$ and nonlinearity for the three regions (compact-support, heavy-tail, and divergent mean).

## VII. CONCLUSION

In conclusion, $q$-statistics and its associated $q$-algebra has been simplified, aligned with the symmetry of the natural numbers, and associated more directly with nonlinear coupling. The translation $q' = 1 - q$ or $1 - q' = q$ which enables the simplification has provided new insights into the relationship between compact-support and heavy-tail $q$-Gaussians. The conjugate transformation

$$\hat{q} = \frac{-2q}{2+q} \quad \text{or} \quad \widehat{q'} = \frac{5-3q'}{3-q'}$$

between these two domains provides a method for extending definitions and theorems in one domain to the other. Using this transformation, the definition for the $q$-Fourier transform has been extended to the compact-support domain:

$$F_q[f(x)](w) = \begin{cases} \int_{\text{supp} f} e_q^{ixw} \otimes_q f(x) dx & -2 < q \le 0, \\ T_{(\hat{q},q)} \circ \int_{\text{supp} f} e_{\hat{q}}^{ixw} \otimes_{\hat{q}} \hat{f}(x) dx & q > 0 \end{cases}$$

In turn the expanded definition of the $q$-Fourier transform should enable the extension of the $q$-Central Limit Theorem to the compact-support domain. The conjugate relationship is also used to define an alternative definition for the $q$-Fourier Transform which provides a transformation between conjugate pairs of $q$-Gaussians. While this conjugate $q$-Fourier Transform requires an additional transformation related to conjugate $q$-exponentials, the resulting pair of functions is closer in form to the original Fourier transform, i.e. wide functions transform to narrow functions and vice-versa.

From a variety of different perspectives, the translated $q$ parameter is shown to be directly proportional to nonlinear coupling. First, the translated definitions for the $q$-entropy sum of independent systems and the $q$-sum have nonlinear terms which are scaled by $q$. Second, the escort probability or coupled probability distribution is expressed in (8) to demonstrate a nonlinear coupling between the statistical states. Finally, the physical parameters for many nonextensive applications are shown to be proportional to $q$ in the compact-support domain or $\hat{q}$ in the heavy-tail domain.

By simplifying the relation to nonlinear analysis it is anticipated that the analytical tools of $q$-statistics will find wider utility in nonlinear information theory. Nonlinear dynamics affect information systems at the device level, where the details of quantum mechanics [25] are increasingly important, at the circuit-board level, where multiplicative noise [23] impacts signal transmission and detection, and at the system-level, where disperse correlations between system components [26] create nonlinear behavior. The analytical tools discussed here provide methods for the modeling and simulation of nonlinear statistical systems in a simple, intuitive form.

## APPENDIX I
### THE TRANSLATED $q$-ALGEBRA EXPRESSIONS

Combinations of the $q$-exponential and $q$-logarithm function form the basis of the $q$-algebra. The $q$-exponential functions, $e_q^x = [1+qx]_+^{\frac{1}{q}}$, can be combined using $q$-addition $\oplus_q$ in the following manner

$$e_q^x e_q^y = [1+qx]^{\frac{1}{q}} [1+qy]^{\frac{1}{q}}$$
$$= [1+q(x+y+qxy)]^{\frac{1}{q}} \tag{57}$$
$$= e_q^{x \oplus_q y}$$

As this example shows, $q$-addition typically combines exponents of a $q$-exponent. The $q$-product typically combines the $q$-exponential functions $f \otimes_q g = [f^q + g^q - 1]_+^{\frac{1}{q}}$. For example,

$$e_q^{x+y} = [1+q(x+y)]^{\frac{1}{q}}$$
$$= [(1+qx)+(1+qy)-1]^{\frac{1}{q}} \tag{58}$$
$$= [e_1^{qx} + e_1^{qy} - 1]^{\frac{1}{q}} = e_q^x \otimes_q e_q^y$$

Combinations of $q$-logarithms, $\ln_q x = \frac{x^q-1}{q}$, follow from these definitions. Thus, the $q$-logarithm can be combined using the following relationships $\ln_q(x \otimes_q y) = \ln_q x + \ln_q y$ and $\ln_q(xy) = \ln_q x \oplus_q \ln_q y$.

To emphasize the benefits of the proposed translation of the $q$ parameter, compare the expressions for three terms and then examine the simplicity of adding n terms with the new expression

original expression:
$$x \oplus_{q'} y \oplus_{q'} z = x + y + z + xy + xz + yz$$
$$- q'(xy + xz + yz) + xyz - 2q'xyz + q'^2 xyz$$
translated expression: (59)
$$x \oplus_q y \oplus_q z = x + y + z + q(xy + xz + yz) + q^2 xyz$$
$$\sum_{i=0}^{N} {}_q x_i = \sum_{i=0}^{N} x_i + q \sum_{i=0}^{N-1} \sum_{j=i+1}^{N} x_i x_j + \ldots q^{N-1} \sum_{i=0}^{1} \prod_{j=i}^{i+N-1} x_j + q^N \prod_{i=0}^{N} x_i$$

The $q$-product expression is also simplified:
Original expression:
$$x \otimes_{q'} y \otimes_{q'} z = \left[ x^{1-q'} + y^{1-q'} + z^{1-q'} - 2 \right]^{\frac{1}{1-q'}}$$
Translated expression: (60)
$$x \otimes_q y \otimes_q z = \left[ x^q + y^q + z^q - 2 \right]^{\frac{1}{q}}$$
$$\prod_{i=1}^{N} {}_q x_i \equiv \left[ x_1^q + x_2^q + \ldots x_N^q - (N-1) \right]^{\frac{1}{q}}$$

The $q$-algebra provides a methodology for concise expression of the relationships between groups of $q$-exponents and $q$-logarithms. For this reason, the connection between $q$-addition and $q$-product is not direct: $x \oplus_q x \oplus_q x \ne 3 \otimes_q x$. Rather the $q$-addition of like $q$-exponents connects with raising of $q$-exponentials to a power: $\exp_q(x \oplus_q x \oplus_q x) = (e_q^x)^3$. And the $q$-product of like $q$-exponentials connects with the multiplication of $q$-exponents: $e_q^x \otimes_q e_q^x \otimes_q e_q^x = e_q^{3x}$. These two expressions are not equal, because raising a $q$-exponential to a power rescales both the exponent and the $q$-parameter

$$\left[ e_q^x \right]^\gamma = [1+qx]^{\frac{1}{q} \cdot \gamma} = \left[ 1 + \left( \frac{q}{\gamma} \right) \gamma x \right]^{\frac{\gamma}{q}} = e_{q/\gamma}^{\gamma x} \tag{61}$$

Thus $\left( e_q^x \right)^3 = e_{q/3}^{3x}$ which does not equal $e_q^{3x}$ unless $q=0$, which recovers the standard exponential function. The $q$-subtraction and $q$-division operations follow in a similar manner and are defined as follows

$$x \ominus_q y = \frac{x-y}{1+qy}.$$
$$x \oslash_q y = \left[ x^q - y^q + 1 \right]^{\frac{1}{q}} \tag{62}$$

Table A.1 summarizes the parameter translation from 1-$q$ to $q$ of the basic expressions of the $q$-algebra.

### ACKNOWLEDGMENT

The authors are grateful for conversations with Constantino Tsallis. P. K. Rastogi provided valuable comments during preparation of the manuscript.

*Table 1:* Examples of the relationship between the entropic index and the nonlinear coupling as defined by (56).

| Application | $q$ and $\alpha$ parameters range of $q$ | Nonlinear Statistical Coupling, $\phi$ | Physical Parameters |
|---|---|---|---|
| Degree of Freedom for Student-T | $q = \dfrac{-2}{\nu+1}$, $\alpha = 2$ <br> $-2 < q \le 0$ | $1/\nu$ | $\nu$ - degree of freedom |
| Correlated anomalous diffusion [27] | $q = \upsilon - 1$, $q > -2$ | Not applicable[*] | $\nu$ - exponent of the nonlinear Fokker-Planck-like equation. |
| Lévy-like anomalous diffusion [27] | $q = \dfrac{-2}{\gamma_L + 1}$, $\alpha = 2$ <br> $-2 < q < -2/3$ | $1/\gamma_L$ | $\gamma_L$ - index of Lévy distribution |
| Multiplicative Noise [23] | $q = \dfrac{-2M}{\tau + M}$, $\alpha = 2$ <br> $-2 < q < 0$ | $M/\tau$ | $M$ – multiplicative noise amplitude <br> $\tau$ - proportional to ratio between deterministic and noise-induced drift |
| Scale-free Networks [27] | $q = \dfrac{-m}{m(3-2r)+1-p-r}$ <br> $\alpha = 2$ <br> $-2 < q < 0$ | $\dfrac{m}{(2m+1)(1-r)-p}$ | $m$ – new links <br> $p$ – prob. of link <br> $r$ – prob. of rewire |
| Logistic Map – sensitivity to initial cond. [27] | $q = \dfrac{\ln 2}{(z-1)\ln \alpha_F(z)}$ <br> $\alpha = 1$ <br> $q \ge 0$ | $\dfrac{\ln 2}{(z-1)\ln \alpha_F(z)}$ | $z$ – exponent of logistic map <br> $\alpha_F(z)$ - Feigenbaum constant |
| Lotka-Volterra Model [27, 28] | $q = 1/d$, $\alpha = 1$ <br> $q > 0$ | $1/d$ | $d$ – dimension of growth |
| Thermostatics with finite bath [24][†] | $q = 1/\beta H$, $\alpha = 1$ <br> $q > 0$ | $1/\beta H$ | $\beta$ - inverse temperature <br> $H$ – total energy of system and bath |

[*] The parameter $\nu$ is defined for all real numbers and therefore all real values of $q$ are appropriate. This application is similar to (3) which is a general solution to a nonlinear equation.
[†] The reference defines $q$ as the negative of the value defined by Tsallis.

*Table VII.1* Comparison of the original and translated *q*-algebra expressions.

| Function Name | Original Parameter, $q'$<br>$q'=1$ is Gaussian | Proposed Translation<br>$q=1-q'$ and $1-q=q'$<br>$q=0$ is Gaussian |
|---|---|---|
| Exponential | $e_{q'}^x \equiv [1+(1-q')x]_+^{\frac{1}{1-q'}}$ | $e_q^x \equiv [1+qx]_+^{\frac{1}{q}}$ |
| Logarithm | $\ln_{q'} x \equiv \dfrac{x^{1-q'}-1}{1-q'}$ | $\ln_q x \equiv \dfrac{x^q-1}{q}$ |
| Escort or Coupled Probability Distribution | $p_i^{q'} \Big/ \sum_{j=1}^W p_j^{q'}$ | $p_i^{1-q} \Big/ \sum_{j=1}^W p_j^{1-q}$ |
| Entropy | $S_{q'} = \dfrac{1-\sum_{i=1}^W p_i^{q'}}{q'-1}$ | $S_q = \dfrac{-1+\sum_{i=1}^W p_i^{1-q}}{q}$ |
| q-Entropy expressed as average of q-surprise | $S_{q'} = \left\langle \ln_{q'} \dfrac{1}{p_i} \right\rangle$ | $S_q = \left\langle \ln_q \dfrac{1}{p_i} \right\rangle$ |
| q-addition | $x \oplus_{q'} y = x+y+(1-q')xy$ | $x \oplus_q y = x+y+qxy$ |
| q-subtraction | $x \ominus_{q'} y = \dfrac{x-y}{1+(1-q')y}$ | $x \ominus_q y = \dfrac{x-y}{1+qy}$ |
| q-product | $x \otimes_{q'} y = [x^{1-q'}+y^{1-q'}-1]_+^{\frac{1}{1-q'}}$ | $x \otimes_q y = [x^q+y^q-1]_+^{\frac{1}{q}}$ |
| q-division | $x \oslash_{q'} y = [x^{1-q'}-y^{1-q'}+1]_+^{\frac{1}{1-q'}}$ | $x \oslash_q y = [x^q-y^q+1]_+^{\frac{1}{q}}$ |
| q-Gaussian sequence<br>$n=0,\pm 1,\pm 2,...$ | $q'_n = z_n(q') = \dfrac{2q'+n(1-q')}{2+n(1-q')}$ | $q_n = z_n(q) = \dfrac{2q}{2+nq}$<br>$=\left[\dfrac{1}{q}+\dfrac{n}{2}\right]^{-1}$ |
| q-alpha sequence for<br>q-alpha-distribution<br>$n=0,\pm 1,\pm 2,...$ | $q'_n = z_n(q') = \dfrac{\alpha q'+n(1-q')}{\alpha+n(1-q')}$ | $q_n = z_n(q) = \dfrac{\alpha q}{\alpha-nq}$<br>$=\left[\dfrac{1}{q}+\dfrac{n}{\alpha}\right]^{-1}$ |
| Conjugate<br>q-Gaussian Dual | $-q'_1 = \dfrac{5-3q'}{3-q'}$ | $-q_1 = \dfrac{-2q}{2+q}$ |
| Conjugate<br>q-exponential Dual | $-q'_2 = \dfrac{3-2q'}{2-q'}$ | $-q_2 = \dfrac{-q}{1+q}$ |
| Additive Duality | $q'_a(q') = 2-q'$ | $q_a(q) = -q$ |
| Multiplicative Inversion | $q'_m(q') = 1/q'$ | $q_m(q) = \dfrac{-q}{1-q} = \dfrac{1}{1-q^{-1}}$ |
| Multiplication & Difference | $q'_{n-1}+\dfrac{1}{q'_{n+1}} = 2$ | $q_{n-1} \cdot q_{n+1} = q_{n-1}-q_{n+1}$ |
| Harmonic Mean[*] | $\dfrac{2}{1-q'} = \dfrac{1}{1-q'_n}+\dfrac{1}{1-q'_{-n}}$ | $\dfrac{2}{q} = \dfrac{1}{q_n}+\dfrac{1}{q_{-n}}$ |

[*]Pajuelo [29] has shown an interesting relationship between the harmonic, arithmetic, and geometric mean which is consistent with the triplet of *q*-values measured for the solar wind [5, 30].

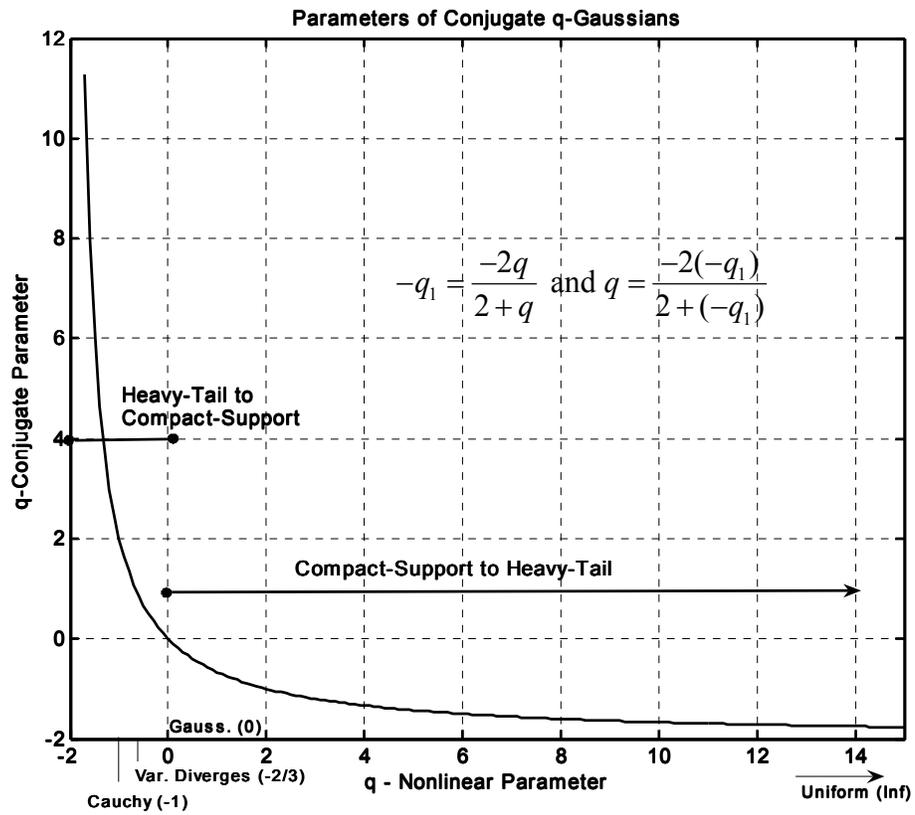

Figure 1: The complementary $q$ parameters for heavy-tail ($-2 < q < 0$) and compact-support ($0 < q < \infty$) $q$-Gaussian functions are related by the self-dual relationship: $-q_1 = \frac{-2q}{2+q}$. This complementary relationship establishes a one-to-one mapping between the two domains of the $q$-Gaussians. The Gaussian function ($q = 0$) is invariant for this mapping.

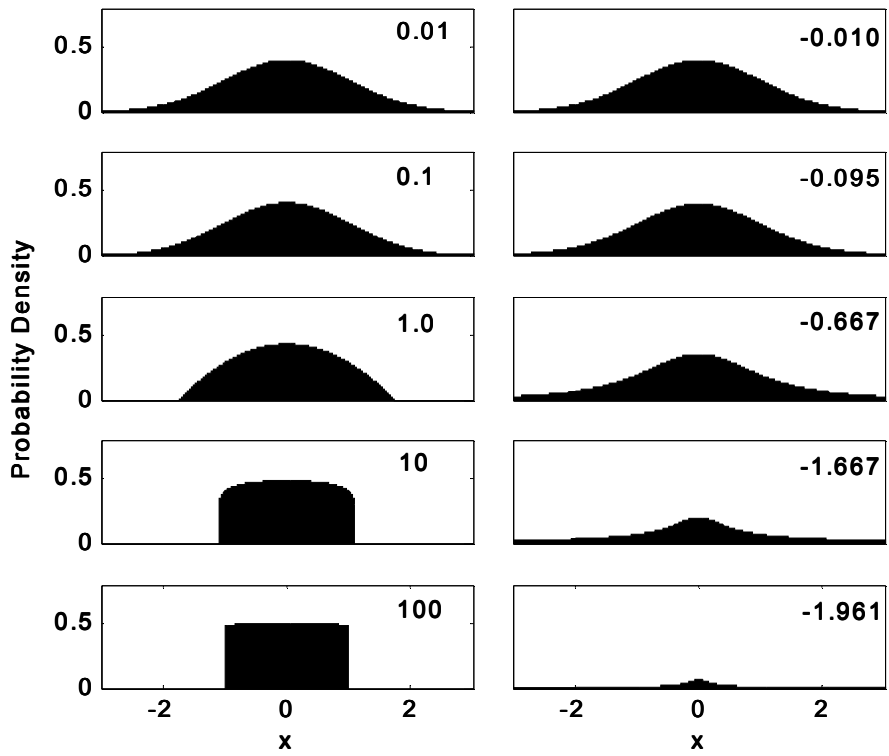

Figure 2: The conjugate pairs of $q$-Gaussian distributions with $\sigma_q^2 = 1$. The value of $q$ is inset in each figure. On the left are the compact-support $q$Gaussians with $q > 0$. On the right are the conjugate heavy-tail distributions with $-2 < q < 0$. The conjugate distribution has $\hat{q} = -q_1 = \frac{-2q}{2+q}$. As $q$ approaches -2 the density is spread infinitesimally to the tails of the distribution.

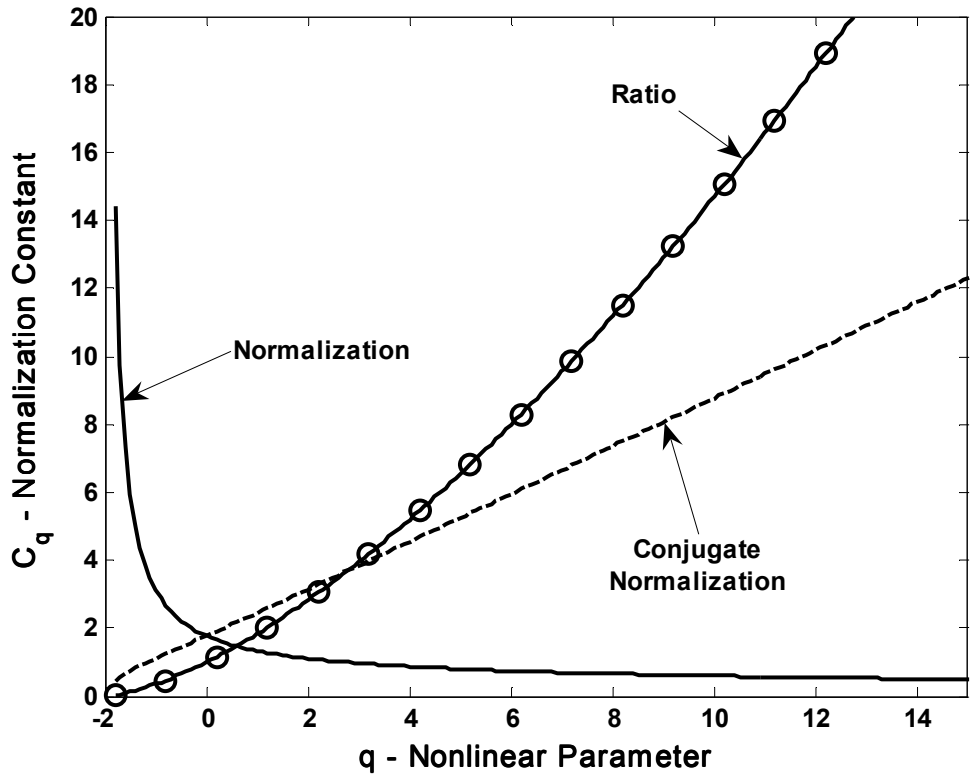

Figure 3: The qGaussian pdf normalization constant, $C_q$ and the conjugate normalization constant, $C_{-q_1}$. The ratio (line) of the conjugate normalization constants is equal to $\frac{C_{-q_1}}{C_q} = \left(\frac{2+q}{2}\right)^{1.5} = \left(\frac{q}{q_1}\right)^{1.5}$ (circles).

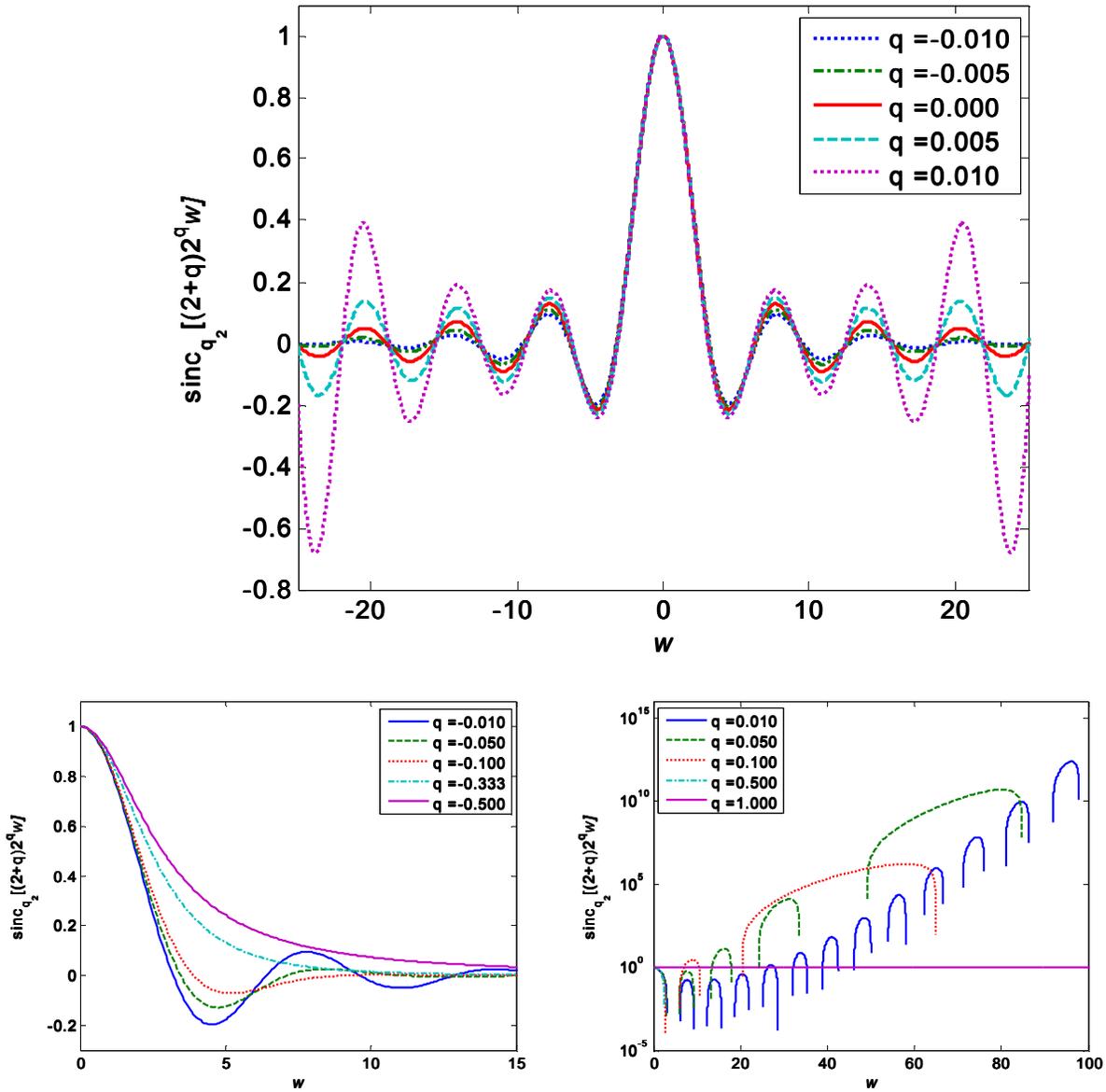

Figure 4: The $q$-Fourier Transform of the uniform distribution is $\text{sinc}_{q_2}[(q+1)2^q w]$. a) Near $q = 0$ the oscillations of a sinc function are evident. Negative values of $q$ dampen the oscillations and positive values of $q$ amplify the oscillations. b) As $q$ varies from -0.01 to -0.04 the oscillations are further dampened. $q = -1/3$ is the critical damping value. c) For $q$ between 0 and 0.1 the oscillations are amplified. A semilog plot is used to show the exponential scale of the oscillations. The gaps are negative values of the functions. For $q$ near 0.5 the function is negative except near $w = 0$. For $q = 1$ the function is one for all values of $w$. For $q > 1$ the function increases monotonically.